\documentclass[a4paper,12pt]{article}

\pdfoutput=1

\usepackage[T1]{fontenc}
\usepackage{soul}
\usepackage[utf8]{inputenc}
\usepackage[english]{babel}
\usepackage{amsmath}
\usepackage{amsthm}
\usepackage{amssymb}
\usepackage{mathrsfs}
\usepackage{bm}
\usepackage{braket}
\usepackage{bbm}
\usepackage{tocloft}
\usepackage{mathtools}
\usepackage{ifthen}
\usepackage{slashed}
\usepackage{graphicx}
\usepackage{epstopdf} 
\usepackage{soul}
\usepackage{xspace}
\usepackage{microtype}
\usepackage{booktabs}
\usepackage{array, tabularx}
\usepackage{chngpage}
\usepackage[english]{varioref}
\usepackage[nottoc]{tocbibind}
\usepackage[numbers,sort&compress]{natbib}
\usepackage{url}
\usepackage{multicol, multirow}
\usepackage{mparhack}
\usepackage{emptypage}
\usepackage{setspace}
\usepackage{dsfont}
\usepackage{rotating}
\usepackage{enumitem}
\usepackage{titlesec}
\usepackage{comment}
\usepackage{tikz,tikz-cd}
\usepackage{xcolor}

\usetikzlibrary{cd,shadings}

\numberwithin{equation}{section}  

\usepackage[left=2.5cm, right=2.5cm, top=2.5cm, bottom=2.5cm]{geometry}
\usepackage[colorlinks=true
,urlcolor=blue
,anchorcolor=blue
,citecolor=blue
,filecolor=blue
,linkcolor=blue
,menucolor=blue
,linktocpage=true
]{hyperref}

\newlength{\bibitemsep}
\newlength{\bibparskip}
\let\oldthebibliography\thebibliography
\renewcommand\thebibliography[1]{%
  \oldthebibliography{#1}%
  \setlength{\parskip}{\bibitemsep}%
  \setlength{\itemsep}{\bibparskip}%
}

\renewcommand{\tilde}{\widetilde}

\renewcommand{\hat}{\widehat}

\DeclareMathOperator{\Tr}{Tr}

\DeclareMathAlphabet{\mathbfsf}{OT1}{cmss}{bx}{n}

\newcommand{\Z}{\mathbb{Z}}

\newcommand{\R}{\mathbb{R}}

\newcommand{\mc}[1]{\mathcal{#1}}

\newcommand{\mf}[1]{\mathfrak{#1}}

\newcommand{\cA}{\mathcal{A}}

\newcommand{\cF}{\mathcal{F}}
\newcommand{\cG}{\mathcal{G}}
\newcommand{\cH}{\mathcal{H}}
\newcommand{\cI}{\mathcal{I}}

\newcommand{\cL}{\mathcal{L}}

\newcommand{\cN}{\mathcal{N}}
\newcommand{\cO}{\mathcal{O}}

\newcommand{\cQ}{\mathcal Q}
\newcommand{\cR}{\mathcal{R}}

\newcommand{\cW}{\mathcal{W}}
\newcommand{\cX}{\mathcal{X}}

\newcommand{\beq}{\begin{equation}}
\newcommand{\eeq}{\end{equation}}

\DeclareMathOperator{\Vol}{\mathrm{Vol}}

\newcommand{\ii}{{\rm i}}
\newcommand{\e}{{\rm e}}

\newcommand{\rd}{{\rm d}}

\newcommand{\vol}{{\rm vol}}
\newcommand{\ph}[1]{\phantom{#1}}

\renewcommand{\j}{\varphi}

\titleformat{\section}{\bfseries}{\thesection.}{4pt}{}
\titlespacing{\section}{0pt}{20pt}{6pt}

\titleformat{\subsection}{\normalfont\itshape}{\thesubsection.}{4pt}{}
\titlespacing{\subsection}{0pt}{15pt}{6pt}

\titleformat{\subsubsection}{\normalfont\itshape}{\thesubsubsection.}{4pt}{}
\titlespacing{\subsubsection}{0pt}{15pt}{6pt}

\titleformat{\paragraph}{\normalfont\itshape}{\theparagraph.}{4pt}{}
\titlespacing{\paragraph}{0pt}{15pt}{6pt}

\allowdisplaybreaks 

\DeclareFontShape{OT1}{cmr}{mx}{n}%
{<->cmr10}{}
\newcommand{\mytitlefont}{\fontseries{mx}\selectfont}
\DeclareMathAlphabet{\titlemath}{OT1}{cmr}{mx}{n}

\newcommand{\bulk}{Y_4}
\newcommand{\bdry}{M_3}
\newcommand{\cutoff}{\delta}
\newcommand{\bulkcutoff}{Y_\delta}
\newcommand{\bdrycutoff}{M_\delta}
\newcommand{\tE}{t_E}
\newcommand{\qE}{q_E}

\begin{document} 

\begin{titlepage}
		
\begin{center}
		
~\\[2cm]
		
{\fontsize{29pt}{0pt} \mytitlefont Magnetic charge and black hole supersymmetric quantum statistical relation}
		
~\\[1.25cm]

Pietro Benetti Genolini$^1$ and Chiara Toldo$^{2,3}$
\hskip1pt

~\\[0.5cm]

{\it $^1$ Department of Mathematics, \\
King’s College London, Strand, London WC2R 2LS, UK}

\vspace{0.3cm}

{\it $^2$ Department of Physics, \\
Jefferson Lab, Harvard University, 17 Oxford Street, Cambridge, MA 02138, USA}

\vspace{0.3cm}

{\it $^3$ Dipartimento di Fisica, \\
Università di Milano, via Celoria 6, 20133 Milano MI, Italy}

~\\[1.25cm]
			
\end{center}

\vspace{2.4cm}
			
\noindent 

We study the thermodynamics in the BPS limit of AdS black holes realizing the topological twist. We use a limiting procedure that allows us to reach the extremal point along a trajectory in the space of supersymmetric Euclidean solutions. We show that on this space we can write a quantum statistical relation, which is well-defined in the BPS limit and relies on imposing a suitable constraint among the chemical potentials, due to supersymmetry and regularity. We stress the importance of this in relating the thermal partition function of the dual field theory to the topologically twisted index.

\vfill 
	
\begin{flushleft}
April 2023
\end{flushleft}
	
\end{titlepage}

\setcounter{tocdepth}{3}
\renewcommand{\cfttoctitlefont}{\large\bfseries}
\renewcommand{\cftsecaftersnum}{.}
\renewcommand{\cftsubsecaftersnum}{.}
\renewcommand{\cftsubsubsecaftersnum}{.}
\renewcommand{\cftdotsep}{6}
\renewcommand\contentsname{\centerline{Contents}}


\medskip

\section{Introduction}
\label{sec:Intro}

The partition function is an important theoretical observable in the study of supersymmetric field theories, and its exact computation via localization on supersymmetry-preserving backgrounds has been a fertile field of research. It is a function of the background fields (the chemical potentials) and the AdS/CFT correspondence relates it, in certain limits, to the renormalized on-shell action of an asymptotically locally anti-de Sitter gravity solution with conformal boundary structure equal to the field theory background. The Legendre transform of the partition function counts the number of states with prescribed charges: correspondingly, in the dual bulk picture in presence of a black hole, the Legendre transform of the on-shell action gives the entropy of the black hole as a function of the charges.

Developing a thermodynamic picture for BPS black holes is challenging, since in this case the length of the Euclidean time coordinate is infinite, so the on-shell action naively diverges.\footnote{Generically, the supersymmetry and extremality conditions do not coincide. Adopting the nomenclature of \cite{Cabo-Bizet:2018ehj}, we denote by ``BPS'' the extremal supersymmetric solutions.} One way to circumvent the issue is performing a specific limiting procedure \cite{Cabo-Bizet:2018ehj}, starting with a family of non-supersymmetric thermal black holes in Lorentzian signature and Wick-rotating to Euclidean signature, hence obtaining solutions that are locally the product of a disc and a sphere. One then imposes supersymmetry, and obtains solutions that are formally not extremal, since the periodicity of the Euclidean time coordinate is finite, but cannot be Wick-rotated back to a Lorentzian well-defined black hole, except for the extremal ones, for which the period of the Euclidean time diverges. The thermodynamics introduced on the family of supersymmetric but not extremal solutions can thus be extended to the BPS extremal ones by taking a well-defined limit, and matches the expectations obtained from the large-$N$ limit of the dual field theory.

This picture, combining the AdS/CFT dictionary \cite{Witten:1998qj} and Euclidean quantum gravity \cite{Gibbons:1976ue}, has been warranted in recent years by an extensive analysis of BPS rotating electrically charged black holes, starting from five dimensions \cite{Hosseini:2017mds, Cabo-Bizet:2018ehj}, whose entropy can be computed by taking the Legendre transform of the large-$N$ limit of the superconformal index in a dual field theory. 

In four dimensions, it is possible to have magnetically charged BPS black holes, in addition to the electric ones found in \cite{Kostelecky:1995ei,Cvetic:2005zi,Hristov:2019mqp} and investigated in \cite{Choi:2018fdc, Cassani:2019mms, Bobev:2019zmz, Nian:2019pxj, Benini:2019dyp, Hosseini:2019and, GonzalezLezcano:2022hcf, Bobev:2022wem, BenettiGenolini:2023rkq}. In contrast to the latter, these magnetic solutions, first found in \cite{Romans:1991nq}, and later on generalized in \cite{Gauntlett:2001qs, Cacciatori:2009iz,DallAgata:2010ejj,Hristov:2010ri,Halmagyi:2013uza,Katmadas:2014faa,Halmagyi:2014qza}, have a well defined static limit, and preserve supersymmetry in a different way, realizing the so-called (partial) ``topological twist''. The presence of the magnetic charge indeed changes the conditions to preserve supersymmetry in the bulk \cite{Hristov:2011ye}, and thus the boundary ones as well.\footnote{There is yet another distinct way to obtain dyonic supersymmetric black holes, by considering C-metric solutions describing accelerating black holes with magnetic charge related to the acceleration parameter \cite{Ferrero:2020twa, Cassani:2021dwa,Ferrero:2021ovq,Faedo:2021nub, Ferrero:2021etw}. In the present paper we restrict our attention to dyonic twisted non-accelerating solutions.} In fact, the partition function of the dual field theory in the supersymmetric background computes the topologically twisted index \cite{Benini:2015noa}, instead of the superconformal index. A lot of progress has been made in relating the black hole entropy to the topologically twisted index (see \cite{Zaffaroni:2019dhb} for a review), yet a careful investigation of the thermodynamics in the BPS limit via the limiting procedure described above is still missing.\footnote{Though a different limiting procedure to define the on-shell action of the BPS black hole by studying perturbations to the supersymmetric solution has been proposed in the literature \cite{Azzurli:2017kxo, Halmagyi:2017hmw, Cabo-Bizet:2017xdr}.} The aim of this letter is to fill in this gap, deriving a supersymmetric quantum statistical relation for topologically twisted dyonic solutions and, at the same time,  elucidating the field theory interpretation of the regularity conditions of the gravity solutions. 

\section{Supersymmetric quantum statistical relation}
\label{sec2}

As mentioned, in order to achieve a well defined BPS limit for black hole thermodynamics, we start from non-supersymmetric non-extremal black holes, and perform a Wick rotation to compute the holographically renormalized Euclidean on-shell action, for which the quantum statistical relation,
\begin{equation}
\label{eq:QuantumStatisticalRelation_Original}
    I = - S + \beta \left( E - \Omega J - \sum_a \Phi_a Q_a \right) \, ,
\end{equation}
holds \cite{Gibbons:1976ue}. In \eqref{eq:QuantumStatisticalRelation_Original}, $I$ is the Euclidean on-shell action, $S$ the entropy, $E$, $J$, $Q_a$ the energy, angular momentum and electric charges, $\beta$ is the inverse temperature, and $\Omega$, $\Phi_a$ the angular velocity and electric potentials. The Wick-rotated solution is topologically the product of a Riemann surface and a disc, and the Killing generator of the horizon $\xi$ vanishes at the origin. We impose supersymmetry, and to ensure global regularity of the supersymmetry spinor $\epsilon$, it must be anti-periodic around the circle generated by $\xi$
\begin{equation}
\label{eq:AntiPeriodic_Spinor}
    \e^{\ii \beta \cL_\xi} \epsilon = -\epsilon \, .
\end{equation}

On the family of supersymmetric solutions, we introduce reduced chemical potentials
\begin{equation}
\label{eq:ReducedChemicalPotentials}
    \tau \equiv \beta \frac{\Omega - \Omega_*}{2\pi\ii} \, , \qquad \j_a \equiv \beta \frac{\Phi_a - \Phi_{a*}}{2\pi\ii} \, , 
\end{equation}
where the subscript $*$ indicates the value of the chemical potential of the BPS black hole. These chemical potentials remain finite everywhere on the family of supersymmetric solutions, so expressing the on-shell action in terms of these allows us to define the on-shell action of BPS black holes via a well-defined limit. Moreover, it allows us to write the quantum statistical relation \eqref{eq:QuantumStatisticalRelation_Original} as
\begin{equation}
\label{eq:QuantumStatisticalRelation_Embryo}
    I = -S - 2\pi\ii \tau J - 2\pi\ii \sum_a \j_a Q_a + \beta \left( E - \Omega_* J - \sum_a \Phi_{a*} Q_a \right)
\end{equation}
As showed in \cite{Cabo-Bizet:2018ehj,Cassani:2019mms} for a number of Wick-rotated electrically charged black holes in various dimensions, the bracketed term vanishes for supersymmetric configurations because it coincides with the supersymmetry condition (the so-called ``BPS bound'') $E = J + \sum_a Q_a$, since $\Omega_* =\Phi_{a*} =1$. Therefore, for these supersymmetric solutions, the quantum statistical relation \eqref{eq:QuantumStatisticalRelation_Original} can be reformulated just in terms of reduced chemical potentials
\begin{equation}
\label{eq:QuantumStatisticalRelation_SUSY}
    I|_{\rm SUSY} = - S - 2\pi\ii \tau J - 2\pi\ii \sum_a \j_a Q_a \, .
\end{equation}
When used in conjunction with the constraint imposed on the chemical potentials by the existence of an anti-periodic Killing spinor, this formula shows that the reduced chemical potentials are the most suitable conjugate variables to the charges in order to express the entropy of the extremal black holes as the constrained Legendre transform of the on-shell action.
Moreover, on the field theory side, the reduced chemical potentials \eqref{eq:ReducedChemicalPotentials} are also the appropriate variables in order to show that the field theory partition function is the index of a supercharge.\footnote{See \cite{Aharony:2021zkr, BenettiGenolini:2023rkq} for a review of the cases of three- and four-dimensional SCFTs.} 

In the following, we show in two families of examples that the modified BPS bound in presence of magnetic charges found in \cite{Hristov:2011ye} still allows us to write the supersymmetric quantum statistical relation \eqref{eq:QuantumStatisticalRelation_SUSY}, and to relate the bulk supersymmetry to the boundary one. Contrary to their electrically charged cousins, the magnetic solutions have a well defined static limit: especially for this reason, we focus our attention on static configurations, with and without scalar profiles, while commenting on the rotating case.

\section{Minimal gauged supergravity}

We begin with solutions of minimal gauged supergravity with the following metric and gauge field \cite{Romans:1991nq, Caldarelli:1998hg} (see appendix \ref{app:Conventions} for our conventions) 
\begin{equation}
\label{eq:Minimal_StaticDyon}
\begin{split}
    \rd s^2 &= - V(r) \, \rd t^2 + \frac{\rd r^2}{V(r)} + r^2 \left( \rd\theta^2 + \sinh^2\theta \, \rd \phi^2 \right) \\
    \cA &= \frac{q}{r} \, \rd t + p \cosh\theta \, \rd\phi + \alpha \, \rd t \,
\end{split}
\end{equation}
where $\alpha$ is a constant and 
\begin{equation}
\label{eq:Minimal_StaticDyon_Vr_v1}
    V(r) = r^2 -1 - \frac{2\eta}{r} + \frac{q^2 + p^2}{r^2} \, .
\end{equation}
We use the coordinates $\theta$, $\phi$ to write a local form of the constant curvature metric on the Riemann surface $\Sigma_g$, normalized so that $\Vol(\Sigma_g) = 4\pi (g-1)$ for $g>1$. 

This solution describes a family of asymptotically locally AdS static dyonic black holes with horizon diffeomorphic to $\Sigma_g$ located at $r_h$, the largest positive root of $V(r)$. The null generator of the horizon is the Killing vector $\xi=\partial_t$. The entropy of the horizon is given by the Bekenstein--Hawking formula
\begin{equation}
\label{eq:Minimal_StaticDyon_Entropy}
    S = \frac{r_h^2}{4G_4} \Vol(\Sigma_g) \, .
\end{equation}
The electrostatic potential of the solution is
\begin{equation}
\label{eq:Minimal_Phi}
    \Phi \equiv \iota_\xi \cA|_{r=r_h} - \iota_\xi \cA|_{r\to \infty} = \frac{q}{r_h} \, .
\end{equation}
We perform a Wick rotation by defining $t = -\ii \tE$. Regularity of the metric allows us to compute the temperature of the horizon as
\begin{equation}
\label{eq:Minimal_StaticDyon_Temperature}
    \beta = \frac{4\pi}{V'(r_h)} \, ,
\end{equation}
and regularity of the gauge field at the horizon requires $\iota_\xi \cA|_{r=r_h} = 0$, so that $\alpha = -q/r_h = -\Phi$.
The resulting solution is a Euclidean metric on $\R^2\times \Sigma_g$, and an imaginary gauge field which can be made real defining $q = \ii \qE$.

The holographic charges of the solutions, computed as in appendix \ref{app:Conventions}, are
\begin{equation}
\label{eq:Minimal_StaticDyon_Charges}
\begin{aligned}
    Q &= \frac{q}{4\pi G_4} \Vol(\Sigma_g) \, , & \qquad P &= - \frac{p}{4\pi G_4} \Vol(\Sigma_g) \, , \\
    E &= \frac{\eta}{4\pi G_4} \Vol(\Sigma_g) \, , &\qquad
    J &= 0 \, .
\end{aligned}
\end{equation}
and the holographically renormalized Euclidean action
\begin{equation}
    I = - \frac{\beta}{8\pi  G_4} \left( r_h^3 - \eta + \frac{q^2-p^2}{r_h} \right) \Vol(\Sigma_g) \, ,
\end{equation}
These quantities satisfy the quantum statistical relation \eqref{eq:QuantumStatisticalRelation_Original} in the form
\begin{equation}
\label{eq:Minimal_StaticDyon_QuantumStatistical_v1}
    I = -S + \beta (E - \Phi Q) \, .
\end{equation}

The black holes \eqref{eq:Minimal_StaticDyon} are supersymmetric provided \cite{Caldarelli:1998hg}
\begin{equation}
\label{eq:Minimal_StaticDyon_SUSY}
    \eta = 0 \, , \qquad p^2 = \frac{1}{4} \, ,
\end{equation}
with electric charge unconstrained. This corresponds to the BPS bound in presence of magnetic charges derived in \cite{Hristov:2011ye}:
\begin{equation}
\label{eq:Minimal_StaticDyon_BPSBound}
    E=0 \, , \qquad P^2 = \left( \frac{\Vol(\Sigma_g)}{8\pi G_4}  \right)^2\,.
\end{equation}
We stress once again that this is \textit{different} from the BPS bound in absence of magnetic charge, namely $E=J+Q$. 

In addition to regularity of the metric and gauge field, for the Wick-rotated supersymmetric solutions we require a globally well-defined Killing spinor $\epsilon$. This imposes that it must be anti-periodic around the circle generated by $\xi$, which in the chosen gauge for $\cA$ imposes
\begin{equation}
\label{eq:Minimal_StaticDyon_Constraint}
    \beta \frac{\Phi}{\pi \ii} = n_0
\end{equation}
with \textit{odd} $n_0$. As advertised, in presence of non-vanishing electric charge the Lorentzian supersymmetric solutions are naked singularities \cite{Caldarelli:1998hg}. Indeed, inserting \eqref{eq:Minimal_StaticDyon_SUSY} in $V(r)$ \eqref{eq:Minimal_StaticDyon_Vr_v1} leads to two branches of solutions with complex $r_h^2$ and generically complex chemical potentials (unless we consider a pure imaginary $q$).

On the other hand, the Lorentzian supersymmetric solution with $q=0$ describes a well-defined extremal black hole. The potentials of the BPS black holes are $\beta_*\to \infty$ and $\Phi_{*}=0$, using which we can introduce the reduced electrostatic potential $\j$ as defined in \eqref{eq:ReducedChemicalPotentials}. Substituting \eqref{eq:Minimal_StaticDyon_Constraint}, this is further constrained to be half-integer, $\j = n_0/2$, in order to have an anti-periodic Killing spinor. Indeed, substituting the value of $r_h$ in \eqref{eq:Minimal_Phi}, for the actual bulk solutions we find $n_0=\pm 1$ depending on the branch.

Inserting the reduced chemical potentials in \eqref{eq:Minimal_StaticDyon_QuantumStatistical_v1} singles out the BPS bound \eqref{eq:Minimal_StaticDyon_BPSBound}, so that on the supersymmetric locus we can write the supersymmetric quantum statistical relation \eqref{eq:QuantumStatisticalRelation_SUSY} in the form
\begin{equation}
\label{eq:Minimal_QuantumStatisticalRelation_SUSY}
    I|_{\rm SUSY} = -S + \beta (\Phi - \Phi_*) Q + \beta E =  - S - 2\pi\ii \j Q \, .
\end{equation}
This is clearly satisfied by the on-shell action of the supersymmetric solutions obtained via direct computation 
\begin{equation}
    I|_{\rm SUSY} = - \frac{1}{8G_4} \Vol(\Sigma_g) = \mp  2\pi \j P \, .
\end{equation}
Here we have rewritten the expression using the reduced electrostatic potential and the magnetic charge $P=\Vol(\Sigma_g)/8\pi G_4$ in order to highlight the fact that we are studying gravity in the grand-canonical ensemble for the electric charge and the canonical ensemble for the magnetic charge. 

The on-shell action $I|_{\rm SUSY}$ does not depend on $\beta$, so the judicious use of variables on the supersymmetric locus allows us to define the on-shell action of the BPS black hole as a limit. To compute its entropy, we perform the Legendre transform of $I|_{\rm SUSY}$ with respect to $\j$, subject to the constraint $\j = \pm 1/2$, which follows directly \eqref{eq:Minimal_QuantumStatisticalRelation_SUSY}
\begin{equation}
    S = \pi (P \mp \ii Q ) \, .
\end{equation}
Notice that the two branches of complex solutions have formally complex conjugate entropy. Imposing reality of the resulting function corresponds to the extremality condition $Q=0$, which leads to $S_* = \pi P$. Thus, restricting the right-hand side of \eqref{eq:Minimal_QuantumStatisticalRelation_SUSY} to the extremal solution, justifies the relation
\begin{equation}
\label{eq:IsusySstar}
    I|_{\rm SUSY} = - S_* \, ,
\end{equation} 
which had been observed via other routes in \cite{Azzurli:2017kxo, Halmagyi:2017hmw, Cabo-Bizet:2017xdr, BenettiGenolini:2019jdz}. Notice that the constraint on the charges imposing reality of the entropy is equivalent to the extremality condition and it is linear, reminiscent of the extremality condition for asymptotically flat black holes (see e.g. \cite{Heydeman:2020hhw, Hristov:2022pmo}), and in contrast to the non-linear relation of asymptotically AdS black holes without a twist \cite{Cassani:2019mms}.

\section{\texorpdfstring{$F = -\ii X^0X^1$}{X0X1} model}

We now extend the considerations of the previous sections to a non-minimal case by looking at the model characterized by prepotential $F=-\ii X^0X^1$. The bosonic sector consists of the metric $\cG$, two $U(1)$ gauge fields $\cA_1$, $\cA_2$, a scalar $X$ and a pseudo-scalar $\chi$ (see appendix \ref{app:Conventions} for more details). The solution we consider is
\begin{equation}
\label{eq:X0X1_BlackHole}
\begin{split}
    \rd s^2 &= - \frac{\mathcal{Q}}{W} \rd t^2 + \frac{W}{\mathcal{Q}} \rd r^2 + W \left( \rd\theta^2 + \sinh^2\theta \, \rd\phi^2 \right) \, , \\
    \cA_1 &= \frac{q_1}{r-\Delta} \rd t + p_1 \, \cosh\theta \, \rd\phi + \alpha_1 \, \rd t \, , \qquad \cA_2 = \frac{q_2}{r+\Delta} \rd t + p_2 \, \cosh \theta \, \rd\phi  + \alpha_2 \, \rd t \, , \\
    X &= \left( \frac{r +\Delta}{r - \Delta } \right)^{\frac{1}{2}} \, , \qquad \chi = 0 \, .
\end{split}
\end{equation}
The functions $\mathcal{Q}$ and $W$ appearing in the metric are
\begin{equation}
    \mathcal{Q} = r^4 + a_2 r^2 -2 a_1 r + a_0 \, , \qquad W = r^2 - \Delta^2
\end{equation}
with
\begin{equation}
\label{param_sol}
\begin{split}
    a_0 &= \frac{q_1^2+q_2^2+p_1^2+p_2^2}{2 } + \Delta^2 \left( 1 + \Delta^2 \right) \, , \\
    a_1 &= - \frac{q_1^2 - p_1^2 - q_2^2 + p_2^2}{4 \Delta} \, , \\
    a_2 &= -1 - 2 \Delta^2 \, .
\end{split}
\end{equation}
The charges are further constrained by\footnote{There also exists a family of static solutions with unconstrained charges \cite{Chow:2013gba}, and rotating extensions thereof \cite{Chow:2013gba,Gnecchi:2013mja}. Euclidean solutions generalizing \eqref{param_sol} have been presented in \cite{Bobev:2020pjk}. However, for the purposes of this paper it is appropriate to work with this solution.}
\begin{equation}
\label{eq:X0X1_Constraint_charges}
    q_1 p_1 - q_2 p_2 = 0 \, .
\end{equation}
 As in the previous case, it describes a family of AlAdS static dyonic black holes with $\Sigma_g$ horizon at $r=r_h$, the largest positive root of $\mc{Q}(r)$, with Killing generator $\xi=\partial_t$, and entropy
\begin{equation}
    S = \frac{W(r_h)}{4G_4} \Vol(\Sigma_g) \, .
\end{equation}
The electrostatic potentials are
\begin{equation}
    \Phi_{1} = \frac{q_1}{r_h - \Delta} \, , \qquad \Phi_{2} = \frac{q_2}{r_h + \Delta} \, .
\end{equation}
Regularity of the Wick-rotated metric and gauge field fixes the temperature and the gauge shifts $\alpha_a$
\begin{equation}
    \beta = \frac{4\pi W(r_h)}{\mathcal{Q}'(r_h)} \, , \qquad \alpha_a = - \Phi_a\,, \qquad   \ a=1,2 \, .
\end{equation}
Using holographic renormalization, we also compute the charges of the solution
\begin{equation}
\begin{aligned}
    Q_a &= \frac{q_a}{8\pi G_4} \Vol(\Sigma_g) \, , &\qquad P_a &= - \frac{p_a}{8\pi G_4} \Vol(\Sigma_g) \, , \\
    E &= \frac{a_1}{4 \pi G_4} \Vol(\Sigma_g) \, , &\qquad J&= 0 \, ,
\end{aligned}    
\end{equation}
and the on-shell Euclidean action
\begin{equation}
\begin{split}
    I &= -\frac{\beta}{8\pi G_4} \left[ r_h^3 - \Delta^2 r_h - a_1 + \frac{1}{2} \frac{q_1^2-p_2^2}{(r_h - \Delta) } + \frac{1}{2} \frac{q_2^2-p_1^2 }{(r_h + \Delta ) } \right] \Vol(\Sigma_g) \, .
\end{split}
\end{equation}
where we need to impose the constraint among charges \eqref{eq:X0X1_Constraint_charges}. Notice that this expression for the on-shell action reduces to the minimal case if we set $\Delta=0$ and $p_1=p_2$.\footnote{For $\Delta=0$ and $p_1=p_2$, $q_1=q_2$ the parameter $a_1$ is unconstrained by the equations of motion, and can be identified with the parameter $\eta$ in \eqref{eq:Minimal_StaticDyon_Vr_v1}} These quantities are related by the quantum statistical relation
\begin{equation}
\label{eq:X0X1_QuantumStatisticalRelation}
    I = -S + \beta \left( E - \Phi_1 Q_1 - \Phi_2 Q_2 \right) \, .
\end{equation}

As showed in appendix \ref{app:Conventions}, the solution is supersymmetric provided 
\begin{equation}
\label{eq:Wick_Rotated_SUSY}
    a_1 = 0 \, , \quad p_a^2 = \frac{1}{4} \qquad \Leftrightarrow \qquad E=0 \, , \quad Q_1 = Q_2 \, , \quad P_a^2 = \left( \frac{\Vol(\Sigma_g)}{16\pi G_4} \right)^2 \, .
\end{equation}
Again, we stress that this BPS bound \cite{Hristov:2011qr} is different from that derived in absence of magnetic charges $E = J + Q_1 + Q_2$ \cite{Cvetic:2005zi}. The Lorentzian solutions are naked singularities, unless we set $q_1 = q_2 = 0$, in which case $\beta_*\to \infty$, and we find BPS black holes with $Q_{1*}=Q_{2*}=0$ and vanishing electrostatic potentials \cite{Cacciatori:2009iz}. 

Performing the Wick rotation we find two branches of supersymmetric solutions labelled by $Q_1=Q_2$ and $\Delta$, with complex chemical potentials, and, following the definitions spelled out in \eqref{eq:ReducedChemicalPotentials}, we define the following reduced electrostatic potentials:
\begin{equation}
    \j_{1} = \pm \frac{1}{2} \left( 1 + \frac{\Delta}{r_h } \right) \, , \quad \j_{2} = \pm \frac{1}{2} \left( 1 - \frac{\Delta}{r_h } \right)
\end{equation}
constrained to satisfy
\begin{equation}
\label{eq:X0X1_Constraint}
    \j_1 + \j_2 = \pm 1 \, ,
\end{equation}
where the different signs correspond to the two branches.
Again, this constraint guarantees that the supersymmetry Killing spinor is anti-periodic around the circle generated by $\xi$, as it should be in order to have disc topology. The introduction of the reduced electrostatic potentials allows us to modify \eqref{eq:X0X1_QuantumStatisticalRelation} into the supersymmetric quantum statistical relation \eqref{eq:QuantumStatisticalRelation_SUSY}
\begin{equation}
\label{eq:X0X1_QuantumStatisticalRelation_SUSY}
    I|_{\rm SUSY} = -S - 2\pi\ii \sum_a \j_a Q_a \, , 
\end{equation}
which indeed holds with 
\begin{equation}
\label{eq:X0X1_ISUSY}
    I|_{\rm SUSY} = - \frac{1}{8G_4} \Vol(\Sigma_g) = \mp 2\pi \left( \j_1 P_2 + \j_2 P_1 \right) \, ,
\end{equation}
choosing $P_a = \Vol(\Sigma_g)/16\pi G_4$.\footnote{Rewriting the action in terms of $\j_a$, $P_a$ is ambiguous. We choose the formula to be consistent with the large-$N$ limit of the partition function of mass-deformed ABJM \cite{Benini:2016rke}.}
Notice that the on-shell action of these solutions is the same as in minimal supergravity, even in presence of the scalar. The constrained Legendre transform leads to $P_1 = P_2$, $Q_1 = Q_2$ and
\begin{equation}
    S = 2\pi ( P_1 \mp \ii Q_1 ) \, ,
\end{equation}
and the reality condition is equivalent to the extremality condition. Moreover, substituting the extremality condition directly in \eqref{eq:X0X1_QuantumStatisticalRelation_SUSY} gives again the relation \eqref{eq:IsusySstar} between the supersymmetric on-shell action and the BPS entropy.

\medskip

The methods described here involve finding supersymmetric saddles of the Euclidean gravitational path integral starting from non-supersymmetric non-extremal Lorentzian black holes.\footnote{On the other hand, there also exist saddles of the path integral that are intrinsically Euclidean, such as the ones found in \cite{Bobev:2020pjk}, that do not have a Lorentzian black hole counterpart and whose on-shell action can matched to a field theory computation.} This explains why we restricted ourselves to considering static black holes, as in the theories we considered
there are no known rotating dyonic BPS solutions with compact horizons. Rotating dyonic twisted BPS solutions with spherical horizon do exist for the $STU$ model \cite{Hristov:2018spe}, which is an extension of the models investigated here, but their non-supersymmetric generalizations are still unknown.

\section{Topologically twisted index}

Finally, we look at the conformal boundary of the supersymmetric solutions described above, and highlight the importance of a non-trivial holonomy of the $R$-symmetry gauge field --- and thus an anti-periodic supercharge --- for the thermal partition function of a three-dimensional $\cN=2$ theory on the supersymmetric background to reproduce the topologically twisted index.

The conformal boundary of the supersymmetric dyonic solutions \eqref{eq:X0X1_BlackHole} with \eqref{eq:Wick_Rotated_SUSY} is a direct product spacetime $S^1\times \Sigma_g$ with two background gauge fields $A_1$, $A_2$, whereas the leading scalar sources $\chi_1$, $\mc{X}$ vanish
\begin{equation}
\label{eq:BdryBackground}
    \rd s^2 = \rd\tE^2 + \rd\theta^2 + \sinh^2\theta \, \rd\phi^2 \, , \qquad A_{a} = p_a \cosh\theta \, \rd\phi + \ii \Phi_a \, \rd\tE \, .
\end{equation}
Additionally, the bulk Killing spinors determine boundary spinors that define three-dimensional $\cN=2$ supersymmetric backgrounds for conformal supergravity \cite{Klare:2012gn}. In fact, expanding the bulk Killing spinor equations near the boundary shows that the boundary spinor is charged under the sum of the two gauge fields, so from the boundary viewpoint, the background gauge field coupling to the $U(1)_R$ $R$-symmetry is 
\begin{equation}
\label{eq:AR_Deformation}
    A^{(R)} = \frac{A_1 + A_2}{2} = \frac{1}{2} \cosh \theta \, \rd\phi + \frac{\ii}{2} (\Phi_{1} + \Phi_{2}) \rd\tE \equiv \frac{1}{2} \cosh \theta \, \rd\phi + \ii \Phi^{(R)} \rd\tE \, .
\end{equation}
The holonomy of $A^{(R)}$ around $S^1_\beta$ is constrained by \eqref{eq:X0X1_Constraint} to be $\Phi^{(R)} = \pm \pi\ii/\beta$, and this guarantees that the boundary spinor is anti-periodic around $S^1_\beta$ (as can be seen by expanding near the boundary the bulk spinor or by solving directly the conformal Killing spinor equation in the background \eqref{eq:BdryBackground}). The difference of the two gauge fields, instead, has the role of a flat background gauge field coupling to a flavor symmetry
\begin{equation}
\label{eq:AF_Deformation}
    A^{(f)} = \frac{A_1 - A_2}{2} = \frac{\ii}{2} \left( \Phi_1 - \Phi_2 \right) \rd \tE = \ii \frac{ q_1 \Delta }{r_h^2 - \Delta^2 } \rd\tE \equiv \ii \Phi^{(f)} \, \rd \tE \, .
\end{equation}

On this background we can formulate a three-dimensional $\cN=2$ field theory while preserving two supercharges with opposite $\mf{u}(1)_R$ charge \cite{Closset:2013vra}. As we quantize the theory on $\Sigma_g$, we find a Hilbert space $\cH_{\Sigma_g}$ with states labelled by the eigenvalues of the Hamiltonian $H$ and the generators of the flavor symmetry $J_f$. The supercharges preserved by the background have anti-commutation relation
\begin{equation}
\label{eq:QQdaggerTTI}
    \{ \cQ, \cQ^\dagger \} = H - 2\pi \sigma_f J_f \, ,
\end{equation}
where $\sigma_f$ is the real mass present in the background vector multiplet coupling to $U(1)_f$. The partition function with thermal boundary conditions on the background \eqref{eq:BdryBackground} has the form
\begin{equation}
\label{eq:ZTTI}
\begin{split}
    Z &= \Tr_{\cH_{\Sigma_g}} \e^{ - \beta H + \beta \Phi^{(R)} R + \beta \Phi^{(f)} J_f } \\
    &= \Tr_{\cH_{\Sigma_g}} (-1)^{R} \e^{ - \beta \{ \cQ, \cQ^\dagger\} + \beta (\Phi^{(f)} - 2\pi \sigma_f ) J_f } \\
    &= \Tr_{\cH_{\Sigma_g}} (-1)^{R} \e^{ - \beta \{ \cQ, \cQ^\dagger\} + 2\pi\ii u_f J_f } \, .
\end{split}
\end{equation}
Here, we massaged the first expression inserting \eqref{eq:QQdaggerTTI}, the constraint on the holonomy of $A^{(R)}$, and the complex combination of the gauge field holonomy and scalar of the flavor background vector multiplet
\begin{equation}
    u_f \equiv - \frac{1}{2\pi}\int_{S^1_\beta} A^{(f)} + \ii \beta \sigma_f \, .
\end{equation}
The manipulations above connect the thermal partition function to a supersymmetric index for the supercharge $\cQ$, graded by the $R$-symmetry, and refined by the flavor symmetry. Crucially, it is the presence of an $R$-symmetry background with non-trivial holonomy that allows a grading while the fermions have anti-periodic boundary conditions. The resulting expression is the topologically twisted index of the field theory \cite{Benini:2015noa, Benini:2016hjo, Closset:2016arn}, which in presence of flavor symmetries generated by $f^\alpha$ is defined as\footnote{If $\Sigma_g \cong S^2$, there is an additional isometry, the states in $\cH_{S^2}$ are labelled by the angular momentum as well, and it is possible to further refine the index \cite{Benini:2015noa}. The bulk dual states would be dyonic rotating solutions with spherical horizons such as \cite{Hristov:2018spe}.}
\begin{equation}
    \cI = \Tr_{\cH_{\Sigma_g}} (-1)^F \e^{ - \beta \{\cQ, \cQ^\dagger \} + 2\pi\ii \j^{(R)} (R + F) + 2\pi\ii \sum_\alpha u_{f^\alpha} J_{f^\alpha} } \, .
\end{equation}
Notice the role of the holonomy of the $R$-symmetry background around $S^1_\beta$, $\j^{(R)}=\beta \Phi^{(R)}/2\pi\ii \in \frac{1}{2}\Z$, which determines the spin structure on $S^1_\beta$ and in turn on the entire $S^1_\beta \times \Sigma_g$ \cite{Closset:2018ghr, Closset:2019hyt}. In the cases just considered, $\j^{(R)}$ is half-integer, which is needed in order to be able to ``fill'' the background and have a bulk gravity dual. On the other hand, if $\j^{(R)}$ is integer, the Killing spinor is periodic around $S^1_\beta$, the topologically twisted index is graded by the fermion number $F$, but this circle cannot bound a disk. This is the analogous relation between the multi-sheeted structure of the superconformal index and the large-$N$ limit \cite{Cassani:2021fyv}.

As a final check, we can look at the large-$N$ limit of the topologically twisted  index of the specific mass-deformed ABJM dual to the solutions considered, since the $F = -\ii X^0X^1$ model is a truncation of $M$-theory on $S^7$ \cite{Cvetic:1999xp, Azizi:2016noi}. In the large-$N$ limit, one finds that in presence of the background gauge fields \eqref{eq:AR_Deformation}, \eqref{eq:AF_Deformation} \cite{Benini:2015eyy, Benini:2016rke}
\begin{equation}
        \log Z_{\rm ABJM} \sim \pm \frac{2\pi\sqrt{2}}{3}N^{\frac{3}{2}} \mf{p}_R\j^{(R)} \, ,
\end{equation}
where $\mf{p}_R=g-1$ is the first Chern number of the $R$-symmetry bundle, provided $\j^{(R)}=\pm 1/2$. This matches the above expectation that the leading large-$N$ behavior requires the anti-periodic spin structure around $S^1_\beta$, and matches \eqref{eq:X0X1_ISUSY}, including the constraint, using the standard dictionary $1/G_4 = 2\sqrt{2}N^{\frac{3}{2}}/3$.

\section*{Acknowledgements}

We are grateful to Davide Cassani, Kiril Hristov, Dario Martelli and Sameer Murthy for helpful discussions and comments on a draft. The work of PBG has been supported by the ERC Consolidator Grant N. 681908, “Quantum black holes: A macroscopic window into the microstructure of gravity”, by the STFC grant ST/P000258/1, and by the Royal Society Grant RSWF/R3/183010.
The work of CT has been supported by the Marie Sklodowska-Curie Global Fellowship (ERC Horizon 2020 Program) SPINBHMICRO-101024314.

\appendix

\section{Appendix}
\label{app:Conventions}

\subsection{Supergravity theories}

The bosonic part of the Lorentzian action of minimal gauged supergravity is
\begin{equation}
\label{eq:Lorentzian_Action_Minimal}
    S = \frac{1}{16\pi G_4} \int \left( \cR + 6 - \cF^2 \right) \vol_{\cG} \, , 
\end{equation}
Here $\cR$ is the Ricci scalar of the metric $\cG$ and $\cA$ is an Abelian gauge field.
A solution is supersymmetric if there is a Dirac spinor $\epsilon$ satisfying
\begin{equation}
\label{eq:Minimal_KSE}
    \left( \nabla_\mu - \ii \cA_\mu + \frac{1}{2} \Gamma_\mu + \frac{\ii}{4} \cF_{\nu\rho} \Gamma^{\nu\rho} \Gamma_\mu \right) \epsilon = 0 \, .
\end{equation}

\medskip

The bosonic part of the Lorentzian action of the $F=- \ii X^0X^1$ model is
\begin{equation}
\label{eq:Lorentzian_Action_X0X1}
\begin{split}
    S = \frac{1}{16\pi G_4} \int_{Y_4} \bigg[ &\cR \, \vol_{\cG} - 2 X^{-2} \, \rd X \wedge * \rd X - \frac{1}{2}X^4 \, \rd \chi \wedge *\rd \chi \\
    & + (4 + X^2 + X^{-2} + \chi^2 {X}^2 ) \, \vol_{\cG} \\
    & - X^{-2} ( \cF_1 \wedge * \cF_1 - \chi X^2 \cF_1 \wedge \cF_1) \\
    & - \frac{1}{X^{-2} + \chi^2 X^2} \left( {\cF}_2 \wedge * {\cF}_2 + \chi X^2 {\cF}_2 \wedge {\cF}_2 \right) \bigg] \, .
\end{split}
\end{equation}
Here $\cG$ is the metric, $\cF_1$ and $\cF_2$ are the curvature of two Abelian gauge fields, $X$ and $\chi$ are a scalar and a pseudo-scalar, respectively. This model reduces to \eqref{eq:Lorentzian_Action_Minimal} upon setting $X=1$, $\chi=0$, $\cF_1 = \cF_2 \equiv \cF$. It is an example of $\cN=2$ gauged supergravity with a unique vector multiplet and prepotential $F=-\ii X^0X^1$ (in terms of the symplectic sections). The model can also be seen as a truncation from  $\cN=4$ $Spin(4)$ gauged supergravity.

A solution is supersymmetric if there is a Dirac spinor $\epsilon$ satisfying \cite{Cacciatori:2008ek, Ferrero:2021ovq}
\begin{equation}
\label{eq:X0X1_KSE}
\begin{split}
    0 &= \bigg[ \nabla_\mu - \frac{\ii}{2} \left( \cA_1 + \cA_2 \right)_\mu + \frac{\ii}{4} X^2 \partial_\mu \chi \Gamma_* + \frac{1}{4}(X + X^{-1})\Gamma_\mu + \frac{\ii}{4} X \chi \Gamma_\mu \Gamma_* \\
    & \qquad + \frac{\ii}{8X} \left( \cF_{1 \nu\rho} + \frac{1 - \ii \chi X^2 \Gamma_*}{\hat{X}^2}\cF_{2 \nu\rho} \right) \Gamma^{\nu\rho} \Gamma_\mu \bigg] \epsilon \, , \\[10pt]
    0 &= \bigg[ \frac{1}{2X^3} \left( \cF_{1 \nu\rho} - \frac{1 - \ii \chi X^2 \Gamma_* }{\hat{X}^2} \cF_{2 \nu\rho} \right) \Gamma^{\nu\rho} - \ii \partial_\mu \left( X^{-2} ( 1 - \ii \chi X^2 \Gamma_* ) \right) \Gamma^\mu \\
    & \qquad - \ii \frac{ 1 - X^{-2} ( 1 - \ii \chi X^2 \Gamma_*) }{X} \bigg] \epsilon
\end{split}
\end{equation}
Here $\hat{X}^2 \equiv X^{-2}+\chi^2 X^2$, and $\Gamma_* \equiv \ii \Gamma_{0123}$.

\bigskip

We consider the solution \eqref{eq:X0X1_BlackHole}, for which we choose the vierbein
\begin{equation}
\begin{aligned}
    \e^0 &=  \sqrt{\frac{\mathcal{Q}}{W}} \, \rd t \, , \qquad \e^1 &= \sqrt{\frac{W}{\mathcal{Q}}} \rd r  \, , &\qquad \e^2 &= \sqrt{W} \, \rd \theta \,  , \qquad \e^3 &= \sqrt{W} \sinh\theta \, \rd\phi \, , &
\end{aligned}
\end{equation}
and the basis 
\begin{equation}\label{Filippo}
\begin{aligned}
    \Gamma_0 &= \begin{pmatrix}
        0 & -\ii \sigma_2 \\ -\ii \sigma_2 & 0
    \end{pmatrix} \, , &\quad
    \Gamma_1 &= \begin{pmatrix}
          \sigma_3 & 0 \\ 0&  \sigma_3 
    \end{pmatrix} \, , \\
    \Gamma_2 &= \begin{pmatrix}
        0 & \ii \sigma_2 \\ -\ii \sigma_2 & 0
    \end{pmatrix} \, , &\quad
    \Gamma_3 &= \begin{pmatrix}
         - \sigma_1 & 0 \\ 0& - \sigma_1 
    \end{pmatrix} \, .
\end{aligned}
\end{equation}
Supersymmetry with nontrivial scalar field profile requires
\begin{equation}
a_1 = 0\,, \quad \qquad p_1^2 = p_2^2 = \frac{1}{4}\,, \quad \qquad q_1 =q_2 \, ,
\end{equation}
and the Killing spinor satisfying \eqref{eq:X0X1_KSE} is (with $p_1=p_2 = -1/2$)
\begin{equation}\label{raveaBK}
    \epsilon = \e^{\frac{\ii\, t}{2} (\alpha_1 +\alpha_2)} \left(\sqrt{f_1(r) - f_2(r)} - \ii \Gamma_0 \sqrt{f_1(r) +f_2(r)} \right) \left( \frac{1-\Gamma_1}{2}\right)\left( \frac{1-\ii\Gamma_2 \Gamma_3}{2}\right) \epsilon_0 
\end{equation}
with
\begin{equation}
f_1(r) = \frac{\sqrt{\left(r^2- \left(\Delta^2+\frac{1}{2}\right)\right)^2+q_1^2}}{\sqrt{r^2-\Delta ^2}}\,, \quad \qquad f_2(r) = \frac{2r^2 -2 \Delta ^2- 1}{2  \sqrt{r^2-\Delta ^2}} \,,
\end{equation}
and $\epsilon_0$ a constant spinor. Notice that for constant scalar field $\Delta =0$ and equal electromagnetic charges $q_1=q_2$, $p_1=p_2$, the solution and the Killing spinor reduce to the ones of \cite{Romans:1991nq}, when adapted to the case of hyperbolic horizons.

As a last remark, we notice that anti-periodicity of the spinor of the Wick-rotated solution \eqref{eq:AntiPeriodic_Spinor} imposes
\begin{equation}
    \beta\frac{\Phi_1 + \Phi_2}{2\pi\ii} = n_0
\end{equation}
with \textit{odd} $n_0$, 
in the ``regular'' gauge where $\iota_\xi \cA_a|_{r=r_h}=0$, which is equivalent to \eqref{eq:X0X1_Constraint}.

\subsection{Holographic renormalization}

We present the formulae for the holographic renormalization and the holographic charges of the $F=-\ii X^0X^1$ model, since the minimal supergravity can be obtained as a subcase.

In an asymptotically locally AdS solution of the model there is a Fefferman--Graham coordinate $z$ near the conformal boundary such that the fields have the expansion
\begin{equation}
\label{eq:FGExpansion}
\begin{split}
    \cG &= \frac{\rd z^2}{z^2} + h_{ij}(x,z) \rd x^i \rd x^j \, , \\
    h_{ij} &= \frac{1}{z^2} \left( g_{ij} + z^2 g^{(2)}_{ij} + z^3 g^{(3)}_{ij} + o ( z^3 ) \right) \\
    \cA_a &= A_a + z A_{a}^{(1)} + z^2 A_a^{(2)} + o(z^2) \, , \\
    X &= 1 + z X_1 + z^2 X_2 + o(z^2) \, , \\
    \chi &= z\chi_1 + z^2 \chi_2 + o(z^3) \, .
\end{split}
\end{equation}
The holographic renormalization of a solution $(\bulk, \cG, \cF_a, X, \chi)$ with conformal boundary $(\bdry, g, F_a, X_1, \chi_1)$ starts by introducing a cutoff $z=\cutoff$ near the conformal boundary, obtaining a space $\bulkcutoff$ with an actual boundary $\bdrycutoff$ that is diffeomorphic to $\bdry$. The outward-pointing unit normal to $\bdrycutoff$ is denoted by $n$, and the induced metric on $\bdrycutoff$ is $h = \cG - n \odot n$, which is used to compute the extrinsic curvature $K_{ij}= \tfrac{1}{2} \cL_{n\sharp}h_{ij}$. In Lorentzian signature, the renormalized on-shell action is defined as
\begin{equation}
    I^{(L)} = \lim_{\cutoff \to \infty} \left( I_{\rm bulk} + I_{\rm GHY} + I_{\rm ct} \right) \, ,
\end{equation}
where $I_{\rm bulk}$ is the action \eqref{eq:Lorentzian_Action_X0X1} computed on-shell on $\bulkcutoff$, and the Gibbons--Hawking--York term and counterterm are
\begin{equation}
    I_{\rm GHY} = \frac{1}{8\pi G_4} \int_{\bdrycutoff} K \, \vol_h \, , \qquad I_{\rm ct} = - \frac{1}{8\pi G_4} \int_{\bdrycutoff} \left( 2\cW + \frac{1}{2} R[h] \right) \vol_h \, , 
\end{equation}
with the superpotential is $\cW = \frac{1}{2}\sqrt{ 2+ X^2 + \hat{X}^2 }$.\footnote{This is the minimal set of counterterms cancelling the divergences and compatible with supersymmetry, as showed in \cite{Freedman:2013oja}. Additional finite counterterms are principle allowed, but they vanish for our solution. Indeed terms cubic in the scalar do not contribute to the on-shell action for the solution at hand \eqref{eq:X0X1_BlackHole}, and the contribution coming from terms linear in the scalars and proportional to the boundary Ricci scalar vanish due to the supersymmetric Ward identities as shown in \cite{Freedman:2013oja, Bobev:2020pjk}.}.  However, the resulting on-shell action, as function of $(g,A_a,X_1, \chi_1)$, cannot be identified with the SCFT generating functional. The boundary SCFT contains scalar operators of dimensions $1$ and $2$: whilst $\chi_1$ is indeed the bulk source of a CFT operator of dimension $2$, $X_1$ is not the source of a dual operator of dimension $1$, but rather its VEV. Therefore, $X_1$ should obey alternate boundary conditions, and we should instead view the variable canonically conjugate to $X_1$ as the bulk source for the SCFT operator of dimension $1$ \cite{Klebanov:1999tb}
\begin{equation}
    \cX = \frac{1}{\sqrt{-g}} \frac{\delta I^{(L)}}{\delta X_1} = - \frac{1}{8\pi G_4} \left( X_1^2 - 2 X_2 + \frac{1}{2} \chi_1^2 \right) \, .
\end{equation}
The correct object to compare to the SCFT generating functional is the Legendre transform of $I$ with respect to $\cX$, that is,
\begin{equation}
    \tilde{I}^{(L)} = I^{(L)} - \int_{\bdry} \cX X_1 \, \vol_g \, ,
\end{equation}
extremized with respect of $X_1$ to obtain a functional of $\cX$.

From this functional we can obtain the one-point functions of the dual operators
\begin{align}
\begin{split}
    \langle \tilde{T}_{ij} \rangle &= - \frac{2}{\sqrt{-g}} \frac{\delta \tilde{I}^{(L)}}{\delta g^{ij}} \\
    &= - \frac{1}{8\pi G_4} \lim_{\cutoff\to 0} \frac{1}{\cutoff} \left[ K_{ij} - K h_{ij} + 2 \cW \, h_{ij} - \left( R_{ij}[h] - \frac{1}{2} R h_{ij} \right)  \right] - \cX X_1 \, g_{ij} \\
    &= \frac{1}{8\pi G_4} \left[ \frac{3}{2}g^{(3)}_{ij} + g_{ij} \left( X_1 \chi_1^2 + \frac{1}{2}\chi_1\chi_2 \right) \right] \, ,
\end{split} \\[5pt]
    \langle \cO_{\Delta=2} \rangle &= \frac{1}{\sqrt{-g}} \frac{\delta \tilde{I}^{(L)}}{\delta \chi_1} = - \frac{1}{8\pi G_4} \left( X_1 \chi_1 + \frac{1}{2} \chi_2 \right) \, , \\[5pt]
    \langle \cO_{\Delta=1} \rangle &= \frac{1}{\sqrt{-g}} \frac{\delta \tilde{I}^{(L)}}{\delta \cX} = - X_1^{(1)} \, , \\
\begin{split}
    \langle j_1^i \rangle &= \frac{1}{\sqrt{-g}} \frac{\delta \tilde{I}^{(L)}}{\delta (A_1^{(1)})_i} = - \frac{1}{8\pi G_4} \lim_{\cutoff\to 0} \frac{1}{\cutoff^3} n_\mu \left( X^{-2} \cF_1 + \chi *_{\cG} \cF_1 \right)^{\mu i} |_{z=\cutoff} \\
    &= \frac{1}{8\pi G_4} (A_1^{(1)})^i \, , 
\end{split} \\[5pt]
\begin{split}
    \langle j_2^i \rangle &= \frac{1}{\sqrt{-g}} \frac{\delta \tilde{I}^{(L)}}{\delta (A_2^{(1)})_i} = - \frac{1}{8\pi G_4} \lim_{\cutoff\to 0} \frac{1}{\cutoff^3} n_\mu \left( \hat{X}^{-2} \cF_2 - \chi *_{\cG} \cF_2 \right)^{\mu i} |_{z=\cutoff} \\
    &= \frac{1}{8\pi G_4} (A_2^{(1)})^i \, .
\end{split}
\end{align}
These quantities satisfy the holographic Ward identities \cite{BenettiGenolini:2020kxj}
\begin{equation}
\begin{split}
\label{eq:HolographicWard}
0 &= \langle \tilde{T}^i_{\ph{i}i} \rangle + 2 \langle \mc{O}_{\Delta=1} \rangle \cX + \langle \mc{O}_{\Delta=2} \rangle \chi_1 \, , \\
0 &= \overline{\nabla}_i \langle j_1^i \rangle = \overline{\nabla}_i \langle j_2^i \rangle \, , \\
0 &= \overline{\nabla}^j \langle \tilde{T}_{ji} \rangle + \langle \mc{O}_{\Delta=1} \rangle \overline{\nabla}_i \cX + \langle \mc{O}_{\Delta=2} \rangle \overline{\nabla}_i \chi_1 + (F_1)_{ij} \langle j_1^{j} \rangle + (F_2)_{ij} \langle j_2^{j} \rangle \, ,
\end{split}
\end{equation}
where $\overline{\nabla}$ is the Levi-Civita connection associated to $g$. The holographic Ward identities \eqref{eq:HolographicWard} guarantee that, if $K$ is a vector generating a symmetry of the boundary, that is, $\cL_K g=0$, $\cL_K A_a=0$, $\cL_K \cX = \cL_K \chi_1 = 0$, the following current is conserved, though not gauge-invariant
\begin{equation}
    J_K^i = \left( \langle \tilde{T}^i_{\ph{i}j}\rangle + \langle j_1^i \rangle (A_1)_j + \langle j_2^i \rangle (A_2)_j\right) K^j \, .
\end{equation}
In order to define the corresponding conserved charge, we need a surface $\Sigma_\infty$ of constant $t$ at the boundary, with induced metric $\gamma$, and the future-directed unit vector normal to the hypersurface of constant $t$ in $\bdry$, denoted by $u$. Then, the conserved charge associated to $K$ is
\begin{equation}
    Q_K = \int_{\Sigma_\infty} u_i J_K^i \, \vol_{\gamma} \, .
\end{equation}
The energy and angular momentum are the holographic charges associated to the Killing vectors $\partial_t$ and $- \partial_\phi$, respectively. As already remarked, their definitions are not gauge-invariant. It is canonical to denote by $E$ the energy measured in the gauge where $\iota_{\partial_t} A = 0$, and by $J$ the angular momentum measured in the gauge where $\iota_{\partial_\phi} A_a = 0$.\footnote{For more details on the gauge choices, see \cite{Ferrero:2020twa, Cassani:2021dwa, BenettiGenolini:2023rkq}.}

The electric charges can also be defined holographically
\begin{equation}
\begin{split}
    Q_1 &= \int_{\Sigma_\infty} u_i \langle j_1^i \rangle = \frac{1}{8\pi G_4} \int_{\Sigma_\infty} ( X^{-2} *_{\cG}\cF_1 - \chi \cF_1) \, , \\ 
    Q_{2} &= \int_{\Sigma_\infty} u_i \langle j_2^i \rangle = \frac{1}{8\pi G_4} \int_{\Sigma_\infty} ( \hat{X}^{-2} *_{\cG}\cF_2 + \chi \cF_2) \, ,
\end{split}
\end{equation}
and they are independent of the choice of cohomologous $\Sigma_\infty$, thanks to Maxwell's equations of motion. The Bianchi identities, on the other hand, guarantee that the magnetic charges are well-defined
\begin{equation}
    P_a = - \frac{1}{8\pi G_4}\int_{\Sigma_\infty} F_a \, .
\end{equation}
When taking the reduction to minimal supergravity, it follows from the definitions that
\begin{equation}
    \langle j \rangle = \frac{1}{\sqrt{-g}} \frac{\delta I^{(L)}}{\delta A^{(1)}} = \langle j_1 \rangle + \langle j_2 \rangle \, , \qquad Q = Q_1 + Q_2 \, , \qquad P = P_1 + P_2 \, . 
\end{equation}
Finally, we should mention that the subtleties due to the presence of scalars are absent in the solutions \eqref{eq:X0X1_BlackHole}, since $\chi\equiv 0$ and $\cX=0$, so $\tilde{I}^{(L)} = I^{(L)}$ for these solutions.

\bibliographystyle{JHEP}
{\small
\bibliography{Bib_BH}}

\end{document}